\documentclass[aps,prb,preprint,groupedaddress,showpacs]{revtex4}
\usepackage{graphicx}

\begin{document}

\title{All-optical octave-broad ultrafast switching of Si woodpile photonic band gap crystals}

\author{Tijmen G. Euser}

\author{Adriaan J. Molenaar}
\affiliation{FOM Institute for Atomic and Molecular Physics, Kruislaan 407, 1098 SJ Amsterdam, The Netherlands}

\author{J. G. Fleming}
\affiliation{Sandia National Laboratories, Albuquerque, NM, USA}

\author{Boris Gralak}
\affiliation{Institut Fresnel, Marseille, France}

\author{Albert Polman}
\affiliation{FOM Institute for Atomic and Molecular Physics,~Kruislaan 407, 1098 SJ  Amsterdam, The Netherlands}

\author{Willem L. Vos}
\email[]{w.l.vos@amolf.nl}
\homepage[]{www.photonicbandgaps.com}
\affiliation{FOM Institute for Atomic and Molecular Physics, Kruislaan 407, 1098 SJ Amsterdam, The Netherlands}
\affiliation{Complex Photonic Systems (COPS), MESA$^+$ Research Institute, University of Twente, The Netherlands}


\date{}

\begin{abstract}
We present ultrafast all-optical switching measurements of Si woodpile photonic band gap crystals. The crystals are spatially homogeneously excited, and probed by measuring reflectivity over an octave in frequency (including the telecom range) as a function of time. After 300 fs, the complete stop band has shifted to higher frequencies as a result of optically excited free carriers. The switched state relaxes quickly with a time constant of 18 ps. We present a quantitative analysis of switched spectra with theory for finite photonic crystals. The induced changes in refractive index are well described by a Drude model with a carrier relaxation time of 10 fs. We briefly discuss possible applications of high-repetition rate switching of photonic crystal cavities.
\end{abstract}

\pacs{42.70.Qs, 42.65.Pc, 42.79.-e}

\maketitle

\section{Introduction\label{Introduction}}

Currently, many efforts are devoted to a novel class of dielectric composites known as photonic crystals \cite{Crete}. Spatially periodic variations of the refractive index commensurate with optical wavelengths cause the photon dispersion relations to organize in bands, analogous to electron bands in solids. Frequency windows known as stop gaps appear in which modes are forbidden for specific propagation directions. Fundamental interest in photonic crystals is spurred by the possibility of a photonic band gap, a frequency range for which no modes exist at all. Tailoring of the photonic density of states by a photonic crystal allows one to control fundamental atom-radiation interactions in solid-state environments \cite{Yablonovitch87,lodahl04}. Additional interest is aroused by the possibility of Anderson localization of light by defects added to photonic band gap crystals \cite{John87}.

Exciting prospects arise when photonic band gap crystals are switched on ultrafast time scales. Switching photonic band gap crystals provides dynamic control over the density of states that would allow the switching-on or -off of light sources in the band gap \cite{Johnson02}. Furthermore, switching would allow the capture or release of photons from photonic band gap cavities \cite{Johnson02}, which is relevant to solid-state slow-light schemes \cite{Yanik04}. The first switching of photonic band gaps has been done on Si inverse opals \cite{euser07}.  Switching the directional properties of photonic crystals also leads to fast changes in the reflectivity, where interesting changes have been reported for 2D photonic crystals \cite{Leonard02,Bristow03,Tan04,Mondia05,Fushman07} and first-order stop bands of 3D opal crystals \cite{Mazur03,Beck05}. Surprisingly, however, there has not been much physical interpretation of ultrafast switching experiments. For instance in Ref. \onlinecite{euser07}, the changes in reflectivity were compared to band structure calculations that pertain to infinitely large crystals.

It is well-known that semiconductors have favorable properties for optical switching, hence they are excellent constituents for switchable photonic materials. Moreover, their elevated refractive indices are highly advantageous to photonic crystals \emph{per se}. Therefore, we present ultrafast all-optical switching experiments on Si woodpile photonic band gap crystals. The free carriers are spatially homogeneously excited to facilitate physical interpretation of the results, as opposed to several inhomogeneous experiments \cite{Leonard02,Mazur03}. Our crystals are probed by measuring reflectivity over broad frequency ranges (including the telecom range) as a function of time. We use theory for finite photonic crystals \cite{Gralak03} to quantitatively interpret ultrafast switching of photonic band gap crystals.

\section{Experimental\label{Experimental}}

\begin{figure}
\includegraphics[width=\linewidth]{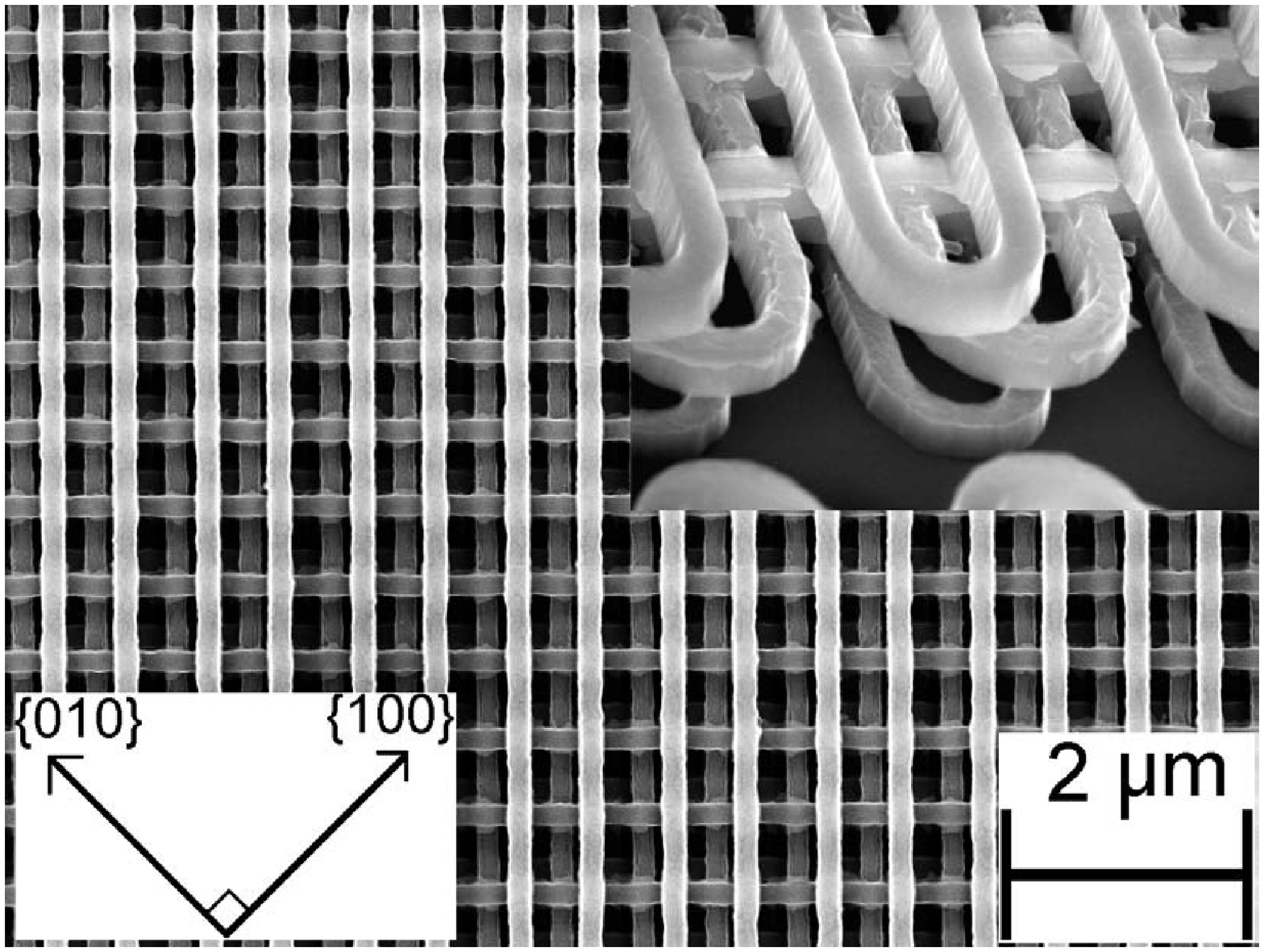}\\
\caption{\label{fig:figure1}High resolution scanning electron micrographs of a (001) surface of a Si woodpile crystal. The average lateral distance between two consecutive rods is 650 $\pm$ 10 nm. The arrows indicate the crystal's (010) and (100) direction. Inset: side view of the crystal. The width and thickness of each rod is 175 $\pm$ 10 nm and 155 $\pm$ 10 nm respectively.}
\end{figure}

The Si woodpile photonic crystals are made using a layer-by-layer approach that allows convenient tuning of the operating wavelengths; here the crystals are designed to have a photonic band gap around the telecommunication wavelength of 1.55 $\mu$m \cite{Ho94,Fleming99}. High resolution scanning electron micrographs of a crystal are shown in Fig.\ \ref{fig:figure1}. The crystals consist of five layers of stacked poly-crystalline Si nanorods that have a refractive index of 3.45 at 1.55 $\mu$m. While each second rod in the crystal is slightly displaced by 50 nm, this periodic perturbation and the resulting superstructure do not affect the photonic band gap region \cite{dedood03}. Our measurements were reproduced on different crystal domains on the same wafer.

A successful optical switching experiment requires an as large as possible switching magnitude, ultrafast time scales, as low as possible induced absorption, as well as good spatial homogeneity \cite{Johnson02}. In Si woodpile photonic crystals, optimal homogeneous switching conditions are obtained for pump frequencies near the two-photon absorption edge of Si at ($\omega$/c)= 5000~cm$^{-1}$ ($\lambda$=~2000~nm) \cite{euser05}. Our setup consists of a regeneratively amplified Ti:Saph laser (Spectra Physics Hurricane) which drives two independently tunable optical parametric amplifiers (OPA, Topas). The OPAs have a continuously tunable output frequency between 3850 and 21050~cm$^{-1}$, with pulse durations of 150~fs and a pulse energy $E_{pulse}$ of at least 20~$\rm{\mu}$J. The probe beam is incident at normal incidence $\theta$= 0$^\circ$, and is focused to a Gaussian spot of 28~$\mu$m FWHM at a small angular divergence NA= 0.02. The E-field of the probe beam is polarized along the (-110) direction of the crystal, that is, perpendicular to the top layer of nanorods. The pump beam is incident at $\theta$= 15$^\circ$, and has a much larger Gaussian focus of 133~$\mu$m FWHM than the probe, providing good lateral spatial homogeneity. We ensure that only the central flat part of the pump focus is probed. The reflectivity was calibrated by referencing to a gold mirror. A versatile measurement scheme was developed to subtract the pump background from the probe signal, and to compensate for possible pulse-to-pulse variations in the output of our laser, see Appendix \ref{Detection scheme}.

\section{Results and Discussion\label{Results and Discussion}}

\begin{figure}
\begin{center}
\includegraphics[width=\linewidth]{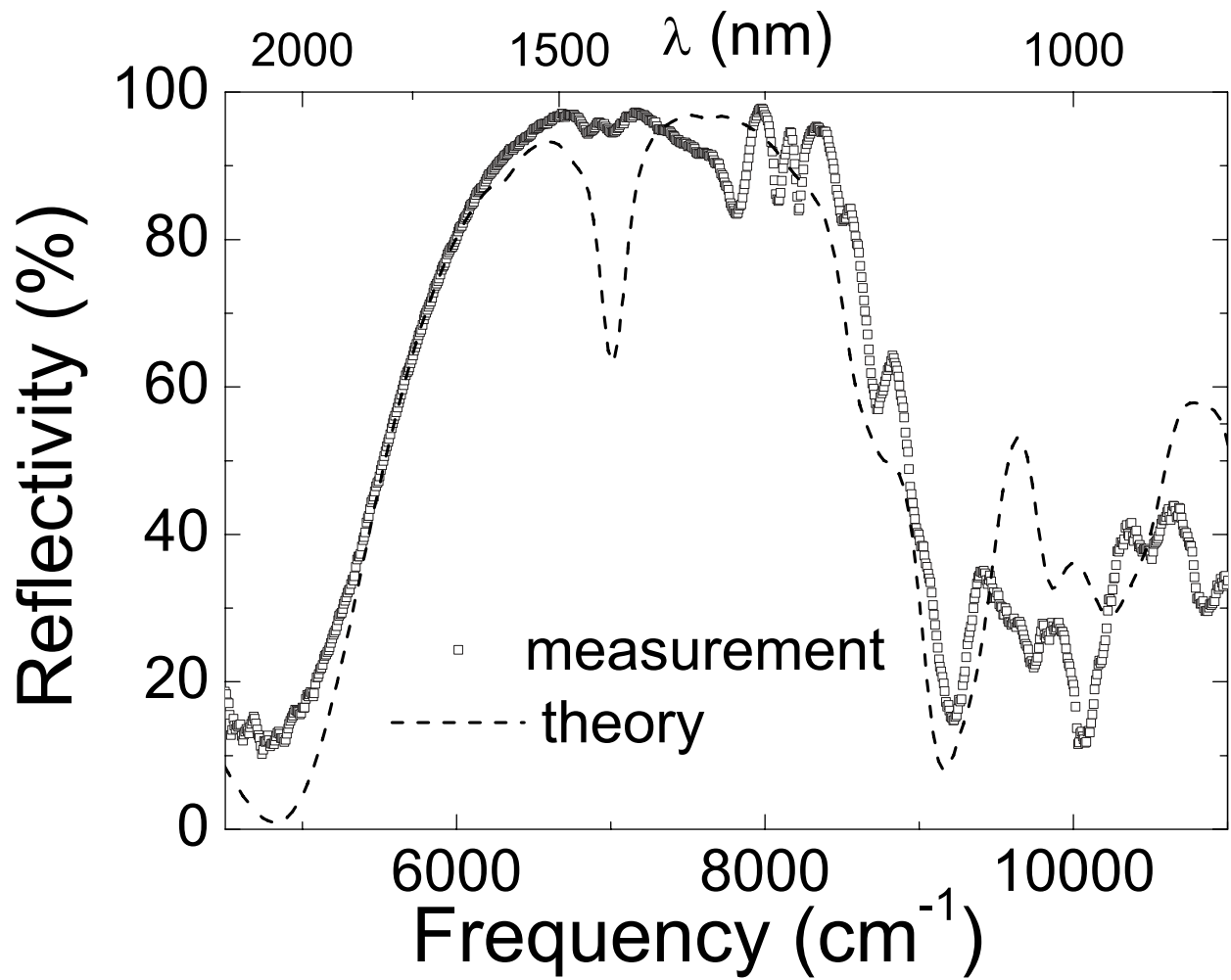}\\
\caption{\label{fig:figure2}(A) Linear unpolarized reflectivity measured in the (001) direction (open squares). A broad stop band with a maximum reflectivity of 95$\%$ appears for frequencies between 5640 cm$^{-1}$ and 8840 cm$^{-1}$. The dashed curve represents an exact modal method calculation for polarized light, that agrees well with the linear measurements in the band gap region.}
\end{center}
\end{figure}

Linear unpolarized reflectivity measurements of the crystal are presented as open squares in Fig.\ \ref{fig:figure2}. The broad stop band from 5600 cm$^{-1}$ to 8800 cm$^{-1}$ corresponds to the $\Gamma$-X stop gap in the band structure, which is part of the 3D photonic band gap of Si woodpile photonic crystals \cite{Gralak03,Ho94}. The large relative width of $\Delta\omega/\omega$= 44$\%$ shows that the crystals interact strongly with the light, in agreement with band gap behavior. While the crystals are relatively thin, the strong photonic interaction strength and the excellent crystal quality result in a high reflectivity of 95$\%$, higher than in bulk Si and than in Si inverse opal photonic structures \cite{euser07,Mazur03}. The dashed curve in Fig.\ \ref{fig:figure2} represents an exact modal method calculation of the reflectivity of the finite crystal in the (001) direction \cite{Gralak03}. The measured Si rod dimensions, the displacements of individual layers, and the superstructure were included in our model. The calculated narrow trough at 7000 cm$^{-1}$ is possibly related to unknown fine details in the superstructure \cite{dedood03}. It is remarkable that the position and width of the stop band in both our measurements and the
theory agree very well, since no parameters were freely adjusted.

\begin{figure}
\begin{center}
\includegraphics[width=\linewidth]{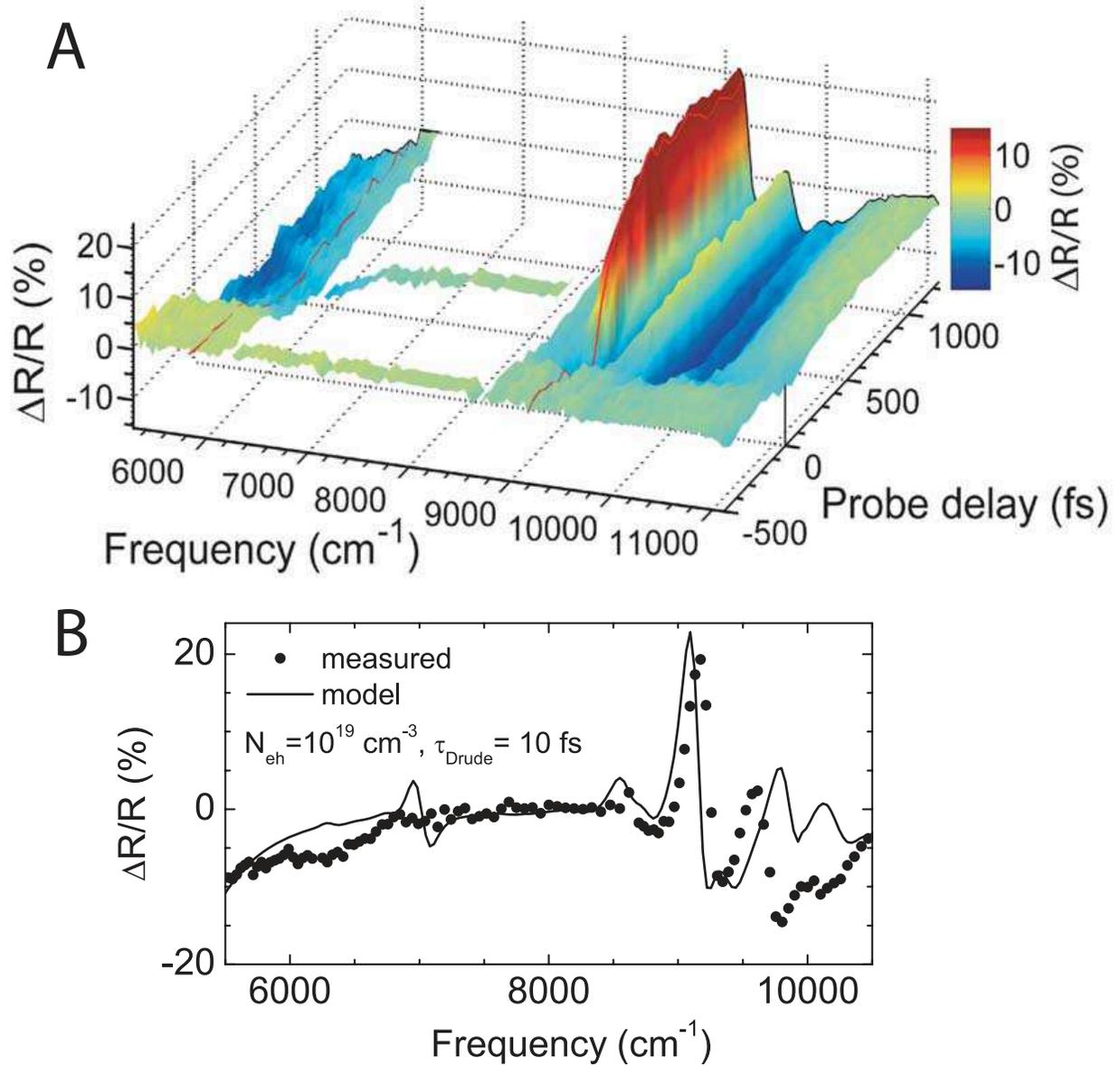}\\
\caption{\label{fig:figure3}(A) (color online) Differential reflectivity versus both probe frequency and probe delay. The pump peak intensity was $I_0$= 17 $\pm$ 1 GWcm$^{-2}$ on the red part, 16 $\pm$ 1 GWcm$^{-2}$ on the central part, and 16 $\pm$ 1 GWcm$^{-2}$ on the blue part of the spectrum. The probe delay was varied in steps of $\Delta$t= 50 fs. The probe wavelength was tuned in $\Delta\lambda$= 10 nm steps in the low, and central range, and in 5 nm steps in the high-frequency range. In the central part of the stop band, $\Delta$R/R($\omega$) was measured at both negative delays and a positive delay of 300 fs. The red curves indicate fixed frequency curves along which extensive delay traces were measured. (B) Measured differential reflectivity changes versus probe frequency, measured at a fixed probe delay of 300 fs (symbols).  The curve indicates differential reflectivity calculated from exact modal method theory that includes the Drude model, and obtained by ratioing to the unswitched calculated spectrum shown in Fig.\ \ref{fig:figure2}. The relative changes at the stop band edges agree
quantitatively with the measured data.}
\end{center}
\end{figure}

Switched spectra were measured with our independently tunable OPAs as a function of probe delay $\tau$ over an octave-broad probe frequency range $\omega_{probe}$.  Our reflectivity measurements were reproduced on different positions on the crystal surface, and were compared to unswitched spectra. The resulting differential reflectivity of the crystal $\Delta$R/R($\tau,\omega_{probe}$) at ultrafast time scales is represented as a three-dimensional surface plot in Fig.\ \ref{fig:figure3}A.  A transient decrease in reflectivity occurs when pump and probe are coincident in time. This effect is attributed to a Kerr effect and non-degenerate two-photon absorption \cite{HardingCLEO07}, and was used to correct our temporal calibration for dispersion in the probe path. At $\tau$= 300 fs after excitation, the reflectivity displays an ultrafast decrease $\Delta$R/R= -7$\%$ at low frequencies (6000 cm$^{-1}$), while at high frequencies near 9170 cm$^{-1}$ (blue gap edge), we observe a strong increase up to $\Delta$R/R= 19$\%$. This distinct dispersive shape in the differential reflectivity is clear evidence of a blue shift of the whole gap. The observation of positive differential reflectivity indicates that the induced absorption remains small. At intermediate frequencies near 7000 cm$^{-1}$, the peak reflectivity of the stop band decreases by less than $\Delta$R/R= -1$\%$, which again signals small induced absorption, as opposed to experiments above the Si-gap where the absorption length is $>$30$\times$ shorter \cite{Mazur03}. At probe frequencies near 9400 cm$^{-1}$, strong variations in $\Delta$R/R with frequency are related to the shift of the superstructure feature (see Fig.\ \ref{fig:figure2}), that are also caused by large refractive index changes of the Si backbone. Compared to bulk Si at similar conditions, the photonic crystal structure results in 10$\times$ enhanced and dispersive reflectivity changes. Our observations lead to the striking conclusion that the entire photonic gap is shifted towards higher frequencies on ultrashort times.

\begin{figure}
\begin{center}
\includegraphics[width=\linewidth]{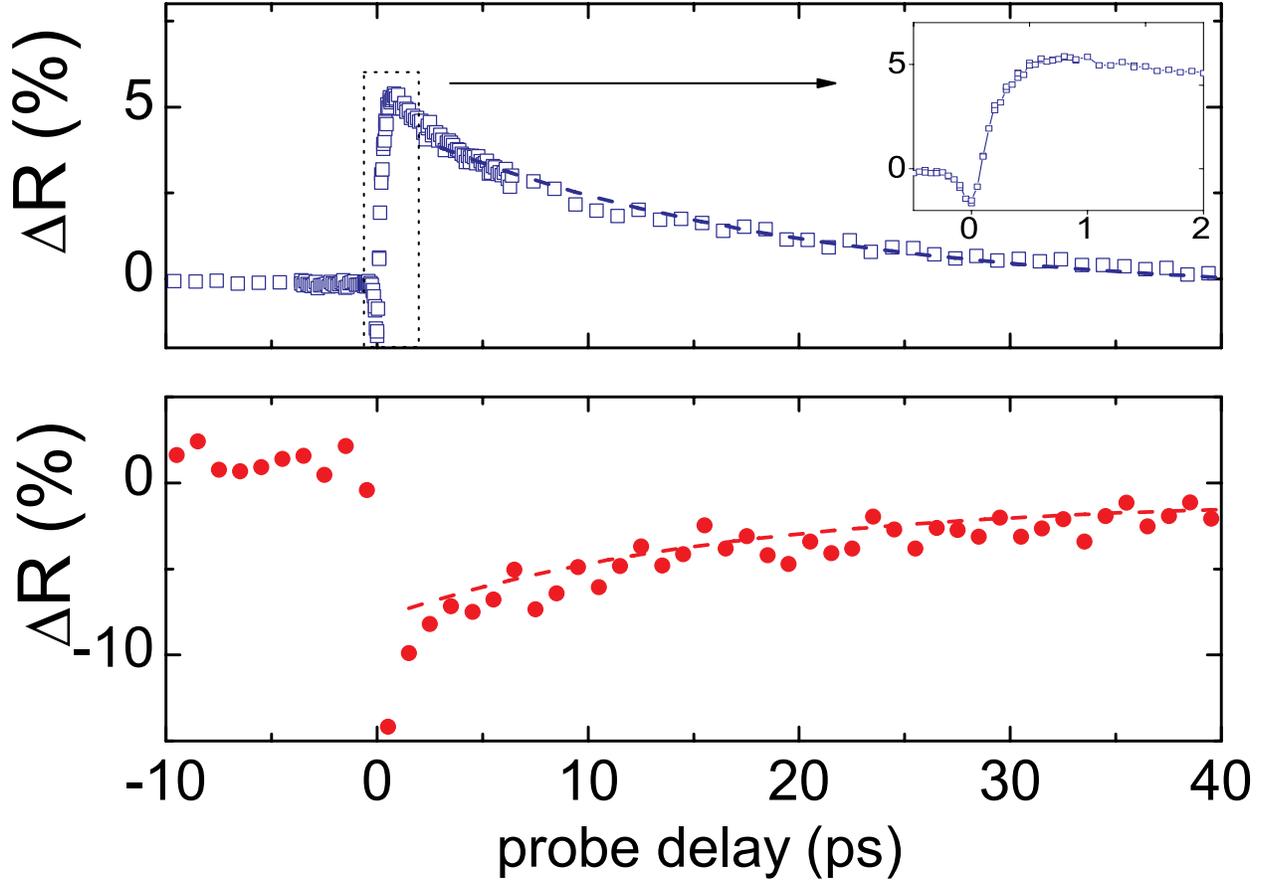}\\
\caption{\label{fig:figure4}(color online) Absolute reflectivity changes versus probe delay at frequency $\omega_{blue}$= 9174 cm$^{-1}$ at the blue edge of the gap (upper panel) and $\omega_{red}$= 5882 cm$^{-1}$ at the red edge (lower panel)of the gap. The pump intensity was $I_0$= 16 $\pm$ 1 GWcm$^{-2}$. The dashed curves are exponential fits with a decay time of 18 ps (upper panel) and 16 ps (lower panel).}
\end{center}
\end{figure}

To study the ultrafast behavior in more detail, we have measured time traces at two characteristic frequencies, namely the red and blue edge of the stop band that are indicated by the red traces in Fig.\ \ref{fig:figure3}A. The time delay curves of the calibrated absolute reflectivity changes $\Delta$R  in Fig.\ \ref{fig:figure4} are measured over an extended time range. At the blue edge, a rapid decrease to $\Delta$R= -1$\%$ appears within 190 fs, followed at 270 fs by an increase to $\Delta$R= 5$\%$ within 500 fs. The subsequent increase is attributed to optically generated free carriers.  The free carrier effect decays exponentially with a decay time of 18 $\pm$ 1 ps. The reflectivity at the red edge decreases by $\Delta$R= -12$\%$ within 1 ps due to the free carrier effect.  After the excitation, the effect on the red edge decays exponentially to $\Delta$R= -1$\%$ with a decay time of 16 $\pm$ 2 ps. The decay times of about 18 ps are much faster than carrier relaxation times in bulk Si, likely since our photonic crystals are made of poly-crystalline silicon, whose lattice defects and grain boundaries act as efficient carrier recombination traps \cite{Yu}.

We compare the switched spectra with theory for finite photonic crystals that includes the complex refractive index of the switched crystal $n'_{Si}(\omega)+in''_{Si}(\omega)$. It appears that the optical properties of excited Si are well described by the Drude model, which is valid for moderate electron densities in the range of our experiments \cite{Sokolowski00}. In our model, the carrier density $N_{eh}$= 2$\times10^{19}$ cm$^{-3}$, and a Drude damping time $\tau_{Drude}$ of 10 fs were deduced by comparing the magnitude of the observed shift at 300 fs and the induced absorption in the stop band to exact modal method theory \cite{Gralak03}. We infer a large maximum change in the real part of the refractive index of the Si backbone $\Delta n_{Si}/n_{Si}$= 2$\%$ at the red edge and $\Delta n_{Si}/n_{Si}$= 0.7$\%$ at the blue edge; such dispersion is typical for Drude effects at probe frequencies above the plasma frequency. From our calculations we obtain the differential reflectivity of the photonic crystal versus probe frequency.  The calculated data (shown as a curve in Fig.\ \ref{fig:figure3}B) agrees well with the measured data at a fixed time delay of $\tau$= 300 fs (filled circles in Fig.\ \ref{fig:figure3}B). Both curves follow the same trend over the full bandwidth of the photonic gap. Both the reflectivity decrease at low frequencies of up to $\Delta$R/R= -10$\%$, as well as the increase $\Delta$R/R= 19 $\%$ at 9000 cm$^{-1}$ are in quantitative agreement with theory.  The small reflectivity decrease of $\Delta$R/R$<$ -1$\%$ in the central part of the peak is also in good agreement with theory.  Small deviations in the calculated reflectivity occur near 7000 cm$^{-1}$ due to a calculated shift of aforementioned narrow trough. Furthermore, the calculations are less accurate for high frequencies above 9000 cm$^{-1}$.

The good agreement between the calculated and measured switched spectra is connected to the notion from photonic bandstructure theory that the band gap for our diamond-like photonic crystals appears in the frequency range of first-order stop gaps \cite{Ho94}. Conversely, for inverse opaline structures, the bandgap is predicted in the range of second-order stop gaps, where observed features are still awaiting a conclusive assignment \cite{Vos00,Galisteo04}. The second order range is also more sensitive to disorder \cite{euser05}. Therefore, we conclude that woodpiles are highly suitable, switchable bandgap crystals.

\begin{figure}
\begin{center}
\includegraphics[width=\linewidth]{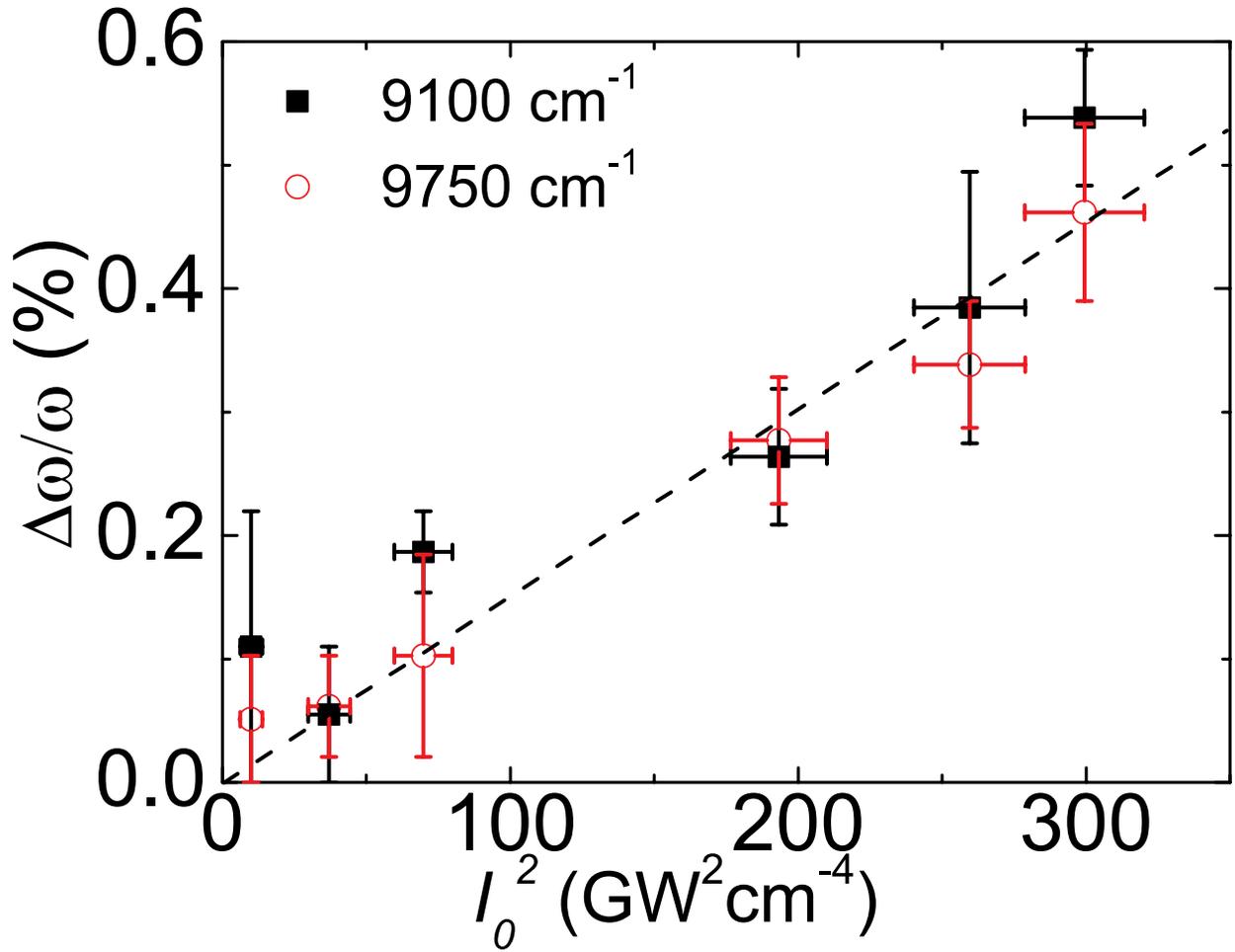}\\
\caption{\label{fig:figure5}(color online) Squares: measured shift $\Delta\omega/\omega$, at $\omega$= 9100 cm$^{-1}$ on the blue edge of the stop band at 1 ps after excitation versus ${I_0}^2$. Diamonds: $\Delta\omega/\omega$ measured at $\omega$= 9750 cm$^{-1}$.  The maximum observed shift is $\Delta\omega/\omega$= 0.54$\%$.  The dashed curve serves to guide the eye.}
\end{center}
\end{figure}

To verify the switching mechanism in the Si backbone of our crystals, we have studied the intensity scaling of the effects. The frequency shift of the
blue edge of the stop band is plotted versus peak pump power squared ${I_0}^2$ in Fig.\ \ref{fig:figure5} for a fixed delay of 1 ps. We have also plotted data for the reflectivity feature at 9750 cm$^{-1}$. Both features shift linearly with the peak pump power squared, which confirms that a two-photon process is indeed the dominant excitation mechanism. From switched reflectivity spectra at 1 ps, we deduce a large maximum shift of the stop band edges of $\Delta\omega_{red}/\omega_{red}$= 2.4$\%$ on the red edge at $\omega_{red}$= 5500 cm$^{-1}$, and at $\omega_{blue}$= 9100 cm$^{-1}$ on the blue edge, the change $\Delta\omega_{blue}/\omega_{blue}$= 0.54$\%$ is smaller, consistent with a Drude dispersion of free carriers. The center position of the gap shifts by 90 cm$^{-1}$ ($\Delta\omega/\omega$= 1.2 $\%$), which is large compared to typical linewidth of quantum dots and of
band gap cavities \cite{Harding07}.

From a comparison of the intensity scaling of the shift of the blue edge (Fig.\ \ref{fig:figure5}) to exact modal method theory, we obtain a two-photon absorption coefficient $\beta$= 60 $\pm$ 15 cmGW$^{-1}$. The corresponding large pump absorption length in the crystal exceeds 230 layers of rods, confirming that two-photon excitation of carriers yields much more homogeneously switched crystals than in one-photon experiments \cite{Mazur03,Mazurenko05}.

Since we have studied photonic crystals made with semiconductor fabrication techniques near the telecom frequency range, it is interesting to generalize our results and consider applications of on-chip ultrafast all-optical switching. Notably important requirements are a considerably reduced pulse energy and a high repetition rate.  A strongly reduced pulse energy is feasible for devices that exploit planar photonic crystal slabs\cite{Noda03} such as modulators wherein a cavity resonance with quality factor Q is switched by one linewidth \cite{Yablonovitch07}.  Since the required refractive index change scales inversely with Q, a small refractive index change $\Delta$n'/n'= 1/Q suffices.  Assuming reported $\lambda^3$-sized 2D photonic crystal cavities with Q= $10^4~$ \cite{Noda03}, $\Delta$n'/n' is 100 times smaller than in our experiments. By pumping such a tiny cavity with diffraction limited pulses from above, free carriers are selectively excited inside the cavity volume only. By choosing a pump frequency of 20000 cm$^{-1}$, which is above the electronic band gap of silicon, a sufficiently high density of free carriers is achieved with low pulse energies of less than 50 fJ, which is within reach of on-chip light sources such as diode lasers. Moreover, the observed decay time of less than 20 ps implies that switching could potentially be repeated at a rate in excess of 25 GHz.  At such high repetition rates, heating of the cavity can be a serious problem. From heat diffusion theory, however, we estimate that the temperature increase in a $\lambda^3$-sized 2D photonic crystal cavity that is pumped with 50 fJ pulses at 25 GHz is less than 10 K, see Appendix \ref{Calculated heating in high rep rate switching}. Therefore, we conclude that ultrafast photonic crystal switching also opens exciting opportunities in device applications.

\section{Summary\label{Summary and Outlook}}
In this Paper, we demonstrate ultrafast switching and recovery of Si woodpile photonic band gap crystals at telecom wavelengths by all-optical free carrier effects. In our switching experiments, we observe a large and ultrafast blue shift of the photonic gap within 300 fs. The switched spectra agree well to theory for finite photonic crystals which includes a Drude description of the free carriers. The good agreement is notably thanks to the spatially homogeneous switching scheme. We demonstrate fast recovery times of 18 ps, which are related to efficient carrier recombination in the polysilicon backbone. We have discussed how such fast switching may be used in applications where high repetition rates are advantageous.

\begin{acknowledgments}
This paper is dedicated to the memory of Jim Fleming. We thank Cock Harteveld and Rob Kemper for technical support, Ad Lagendijk, Dimitry Mazurenko, and Tom Savels for discussions, and Philip Harding for experimental help. This work is part of the research program of the "Stichting voor Fundamenteel Onderzoek der Materie" (FOM), which is supported by the "Nederlandse Organisatie voor Wetenschappelijk Onderzoek" (NWO). WLV thanks NWO-Vici and STW/NanoNed.
\end{acknowledgments}

\appendix

\section{Detection scheme\label{Detection scheme}}
The intensity of each pump and probe pulse is monitored by two photodiodes, and the reflectivity signal by a third photodiode. A boxcar averager holds the short output pulses of each detector for 1 ms, allowing simultaneous acquisition of separate pulse events of all three channels. Both pump and probe beam pass through a chopper that is synchronized to the repetition rate of the laser. The alignment of the two beams on the chopper blade is such that in one sequence of four consecutive laser pulses, both pumped reflectivity, linear reflectivity, and two background measurements are collected. The pump-probe delay was set by a 40 cm long delay line with a resolution corresponding to $\tau$= 10 fs. At each frequency-delay setting, 4x250 pulse events were stored to increase the signal-to-noise ratio. From the resulting large data set, the background was subtracted, and each reflected signal was referenced to its proper monitor signal to compensate for intensity variations in the laser output.

\section{Calculated heating in high rep rate switching\label{Calculated heating in high rep rate switching}}

We consider an  instantaneous point source of energy $E_{pump}$= 50 fJ which is released in a photonic crystal cavity whose thermal properties are assumed to be similar to bulk silicon. The temperature history $\theta_{cav}(r,t)$ of the cavity as a result of this single pump pulse at a distance $r$ from the source is described by standard diffusion theory \cite{Bejan}:

\begin{equation}
\theta_{cav}(r,t)= \frac{E_{pump}}{8\rho c_p(\pi \alpha t)^{3/2}}exp{\Bigg(}-\frac{r^2}{4\alpha t}{\Bigg)}, \label{eq1}
\end{equation}

where  $\rho$= 2330 [kg/m$^3$] the density, $\alpha$= 0.94 [cm$^2$/s] the diffusion constant, and $c_p$= 703 [J/kgK] the heat capacity of silicon.

We now consider the situation in which a continuous series of pulses with energy $E_{pump}$ is released into the cavity at a repetition rate of 25 GHz. The time between two pulses is $\Delta$t= 40 ps. After N pulses, at time t= N$\Delta$t, the temperature distribution is given by:

\begin{equation}
\theta_{cav}(r,t)= \theta_0+\theta_1+\theta_2+ \cdots + \theta_N.
\label{eq2}
\end{equation}

To find the equilibrium temperature in the sample after many pulses, we evaluate Eq. \ref{eq2} at a time t=(N+1)$\Delta$ t, one cycle after the final pulse, and take the limit of the number of pulses going to infinity:

\begin{equation}
\theta(r,t)= \lim_{N\to\infty} \sum_{a=0}^N \frac{E_{pump}}{8\rho c_p(\pi \alpha((N+1) \Delta t-a\Delta t))^{3/2}}exp{\Bigg(}-\frac{r^2}{4\alpha((N+1) \Delta  t -a\Delta t)}{\Bigg)}.
\label{eq3}
\end{equation}

In the center of the cavity, at r=0, Eq. \ref{eq3} reduces to:

\begin{equation}
\theta(r,t)= \lim_{N\to\infty} \sum_{a=0}^N \frac{E_{pump}}{8\rho c_p(\pi \alpha((N+1) \Delta t-a\Delta t))^{3/2}}, \label{eq4}
\end{equation}

which can be simplified by bringing all constant pre factors out of the summation:

\begin{equation}
\theta(r,t)=  \frac{E_{pump}}{8\rho c_p(\pi \alpha\Delta t)^{3/2}} {\Bigg(}\lim_{N\to\infty}\sum_{a=0}^N \frac{1}{((N+1)-a)^{3/2}}{\Bigg)}, \label{eq5}
\end{equation}

The last factor of Eq. \ref{eq5} is the well-known Riemann Zeta function $\zeta$(s) \cite{Abramowitz}. In our case, where s= 3/2, $\zeta$(s) evaluates to $\zeta$(3/2)$\approx$ 2.612. The fixed prefactor (see Eq. \ref{eq5}) is equal to 2.972, yielding a temperature increase in the cavity of only 2.972$\times$2.612= 7.8 K.  We conclude that from the point of heat accumulation, the switching of a photonic crystal cavities with a continuous pulse train of weak 50 fJ pulses with an elevated repetition rate of 25 GHz seems perfectly feasible, and merits a study in its own.

\newpage

\end{document}